\documentclass[]{aastex631}

\shorttitle{Geometric albedos at optical wavelengths}
\shortauthors{Mallonn et al.}

\graphicspath{{./}{figures/}}

\begin{document}

\title{Geometric albedos at short optical wavelengths for the hot Jupiters WASP-43b, WASP-103b, and TrES-3b}


\author{Matthias Mallonn}
\affiliation{Leibniz-Institut f\"{u}r Astrophysik Potsdam, An der Sternwarte 16, D-14482 Potsdam, Germany \\
}

\author{Enrique Herrero}
\affiliation{Institut d’Estudis Espacials de Catalunya (IEEC), C/ Gran Capit\'{a} 2-4, 08034 Barcelona, Spain}

\author{Carolina von Essen}
\affiliation{Stellar Astrophysics Centre, Department of Physics and Astronomy, Aarhus University, Ny Munkegade 120, DK-8000 Aarhus C, Denmark}
\affiliation{Astronomical Observatory, Institute of Theoretical Physics and Astronomy, Vilnius University, Sauletekio av. 3, 10257, Vilnius, Lithuania}

\begin{abstract}

The largest and most close-in exoplanets would reflect enough star light to enable its ground-based photometric detection under the condition of a high to moderate albedo. We present the results of an observing campaign of secondary eclipse light curves of three of the most suitable exoplanet targets, WASP-43b, WASP-103b, and TrES-3b. The observations were conducted with meter-sized telescopes in the blue optical broadband filters Johnson B and Johnson V. We do not detect a photometric dimming at the moment of the eclipse, and derive a best-fit eclipse depth by an injection-recovery test. These depth values are then used to infer low geometric albedos ranging from zero to 0.18 with an uncertainty of 0.12 or better in most cases. This work illustrates the potential of ground-based telescopes to provide wavelength-resolved reflection properties of selected exoplanets even at short optical wavelengths, which otherwise are only accessible by the Hubble Space Telescope.

\end{abstract}



\section{Introduction} \label{sec:intro}

The optical reflection properties of close-in gas giant planets outside our solar system have almost exclusively been measured with space-based telescopes. Wavelength-resolved information below 600~nm exist only for very few, selected cases despite the distinctive power of reflection spectra to reveal day-side clouds and other atmospheric properties \citep{Mayorga2019,Adams2022}. In \cite{Mallonn2019}, we used ground-based observations to derive upper limits for the geometrical albedos in the wavelength transition region between the optical and the near-infrared.
For a given value of geometric albedo, the exoplanets with the largest radius and the least distance to their host star produce the largest photometric dip in their secondary eclipse light curve. As already presented in Table~5 of \cite{Mallonn2019}, we ranked the known exoplanets for their potential eclipse depth, and run an observing campaign for several of the most favorable targets. Here we present the results for the three targets WASP-43b, WASP-103b, and TrES-3b, which are estimated to show an eclipse depth of 0.45~ppt, 0.3~ppt, and 0.23~ppt, respectively, under the assumption of a geometric albedo of 0.3. Our observations present the first ground-based measurements of exoplanet geometric albedos at blue visible wavelengths, see \cite{Hooton2018} for a similar approach at UV wavelengths.

\section{Observations and data reduction}
\label{sec:obs}
In this work, we observed and analyzed 69 photometric light curves, which each fully cover the predicted time interval of the exoplanet secondary eclipse event. The majority of data have been obtained with the robotic 1.2m STELLA telescope and its wide-field imager WiFSIP \citep{Strassmeier2004}. Additional light curves have been observed with the 0.8m Joan Oró telescope (TJO) of the Montsec Observatory and its imaging instrument MEIA2. For WASP-43b, we obtained 14 and 14 light curves in the filters Johnson B and Johnson V, respectively, for WASP-103b seven and five light curves, and for TrES-3b 22 and seven light curves. The data sets show a typical point-to-point scatter of 1.0 to 1.5~mmag in a cadence of 90 to 120 seconds. Only the observations of WASP-43 in the Johnson B filter show a higher scatter of 2.0 to 3.0~mmag due to the faintness of this late K dwarf at short wavelengths. A summary of the observations is provided in Table~\ref{tab_obs}. The data reduction followed the procedure of previous exoplanet time-series photometry with the same instruments \citep{Mallonn2015,Mallonn2019,Mallonn2022}.

\begin{table*}
\tiny
\caption{Overview of observations analyzed in this work. The columns provide the observing date, the number of the observed individual data points, the exposure time, the observing cadence, the dispersion of the data points as root-mean-square (rms) of the observations after a detrending function, and the $\beta$ factor \citep{Mallonn2019}.}
\label{tab_obs}
\begin{center}
\begin{tabular}{llccccccc}
\hline
\hline
\noalign{\smallskip}
Object & Date & Observatory & Filter & $N_{\mathrm{data}}$ &  $t_{\mathrm{exp}}$ (s) & Cadence (s) &   rms (mmag) &  $\beta$ \\
\hline
\noalign{\smallskip}
WASP-43b  & 07.03.20 & STELLA  & B &  92  &   90  &   117  &   3.30  &  1.00  \\ 
WASP-43b  & 12.03.20 & STELLA  & B &  93  &   90  &   116  &   2.53  &  1.01  \\ 
WASP-43b  & 16.03.20 & STELLA  & B &  92  &   90  &   116  &   2.23  &  1.13  \\ 
WASP-43b  & 21.03.20 & STELLA  & B &  92  &   90  &   116  &   1.64  &  1.00  \\ 
WASP-43b  & 29.03.20 & STELLA  & B &  91  &   90  &   117  &   2.14  &  1.00  \\ 
WASP-43b  & 25.04.20 & STELLA  & B &  93  &   90  &   116  &   1.89  &  1.00  \\ 
WASP-43b  & 29.04.20 & STELLA  & B &  89  &   90  &   116  &   3.74  &  1.00  \\ 
WASP-43b  & 08.05.20 & STELLA  & B &  81  &   90  &   117  &   1.90  &  1.03  \\ 
WASP-43b  & 28.01.21 & STELLA  & B &  93  &   90  &   117  &   2.79  &  1.00  \\ 
WASP-43b  & 15.02.21 & STELLA  & B &  72  &   90  &   117  &   1.79  &  1.00  \\ 
WASP-43b  & 19.02.21 & STELLA  & B &  93  &   90  &   117  &   2.42  &  1.17  \\ 
WASP-43b  & 23.02.21 & STELLA  & B &  93  &   90  &   117  &   2.83  &  1.00  \\ 
WASP-43b  & 24.02.21 & STELLA  & B &  93  &   90  &   117  &   3.20  &  1.01  \\ 
WASP-43b  & 09.03.21 & STELLA  & B &  91  &   90  &   117  &   2.83  &  1.00  \\ 
\hline
\noalign{\smallskip}
WASP-43b  & 20.03.19 & STELLA  & V &  68  &   80  &   107  &   1.20  &  1.44  \\
WASP-43b  & 10.12.19 & STELLA  & V & 101  &   80  &   107  &   1.05  &  1.23  \\
WASP-43b  & 19.12.19 & STELLA  & V & 101  &   80  &   106  &   1.68  &  1.51  \\
WASP-43b  & 10.01.20 & STELLA  & V & 102  &   80  &   107  &   1.84  &  1.00  \\
WASP-43b  & 27.01.20 & STELLA  & V &  83  &   80  &   107  &   0.95  &  1.04  \\
WASP-43b  & 28.01.20 & STELLA  & V &  95  &   80  &   107  &   1.01  &  1.00  \\
WASP-43b  & 18.03.21 & STELLA  & V &  93  &   90  &   116  &   1.24  &  1.00  \\
WASP-43b  & 22.03.21 & STELLA  & V &  92  &   90  &   116  &   1.05  &  1.21  \\
WASP-43b  & 31.03.21 & STELLA  & V &  92  &   90  &   117  &   1.62  &  1.31  \\
WASP-43b  & 01.04.22 & STELLA  & V &  93  &   90  &   116  &   1.53  &  1.00  \\
WASP-43b  & 10.04.22 & STELLA  & V &  92  &   90  &   116  &   1.14  &  1.00  \\
WASP-43b  & 14.04.22 & STELLA  & V &  73  &   90  &   116  &   1.45  &  1.30  \\
WASP-43b  & 02.05.22 & STELLA  & V &  93  &   90  &   117  &   1.07  &  1.00  \\
WASP-43b  & 15.05.22 & STELLA  & V &  77  &   90  &   116  &   1.55  &  1.06  \\
\hline
\noalign{\smallskip}
WASP-103b & 24.05.19 & STELLA  & B & 225  &   60  &    86  &   1.44  &  1.07  \\
WASP-103b & 25.05.19 & STELLA  & B & 229  &   60  &    86  &   1.42  &  1.14  \\
WASP-103b & 26.05.19 & STELLA  & B & 228  &   60  &    86  &   1.56  &  1.36  \\
WASP-103b & 06.05.20 & STELLA  & B & 228  &   60  &    86  &   2.17  &  1.09  \\
WASP-103b & 19.05.20 & STELLA  & B & 223  &   60  &    87  &   1.55  &  1.10  \\
WASP-103b & 04.07.21 & STELLA  & B & 210  &   60  &    87  &   1.44  &  1.00  \\
WASP-103b & 20.03.22 & STELLA  & B & 207  &   60  &    86  &   1.36  &  1.00  \\
\hline
\noalign{\smallskip}
WASP-103b & 13.05.19 & TJO     & V & 274  &   60  &    67  &   1.94  &  1.62  \\
WASP-103b & 22.03.19 & STELLA  & V & 201  &   60  &    86  &   1.35  &  1.00  \\
WASP-103b & 29.04.19 & STELLA  & V & 224  &   60  &    86  &   0.99  &  1.47  \\
WASP-103b & 11.05.19 & STELLA  & V & 194  &   60  &    86  &   1.22  &  1.25  \\
WASP-103b & 13.05.19 & STELLA  & V & 212  &   60  &    86  &   1.15  &  1.50  \\
\hline
\noalign{\smallskip}
TrES-3b   & 09.07.19 & STELLA  & B & 101  &   80  &   106  &   1.42  &  1.00  \\
TrES-3b   & 13.07.19 & STELLA  & B &  69  &   80  &   106  &   2.31  &  1.14  \\
TrES-3b   & 26.07.19 & STELLA  & B &  84  &   80  &   106  &   1.35  &  1.00  \\
TrES-3b   & 30.07.19 & STELLA  & B & 101  &   80  &   107  &   1.09  &  1.00  \\
TrES-3b   & 03.08.19 & STELLA  & B & 102  &   80  &   106  &   0.94  &  1.04  \\
TrES-3b   & 16.08.19 & STELLA  & B & 102  &   80  &   106  &   1.86  &  1.00  \\
TrES-3b   & 20.08.19 & STELLA  & B &  99  &   80  &   106  &   1.09  &  1.00  \\
TrES-3b   & 02.09.19 & STELLA  & B & 100  &   80  &   106  &   1.69  &  1.00  \\
TrES-3b   & 06.09.19 & STELLA  & B & 101  &   80  &   106  &   0.96  &  1.13  \\
TrES-3b   & 21.03.20 & STELLA  & B &  66  &   80  &   106  &   1.37  &  1.20  \\
TrES-3b   & 07.05.20 & STELLA  & B & 100  &   80  &   106  &   1.47  &  1.00  \\
TrES-3b   & 01.06.20 & STELLA  & B &  92  &   80  &   106  &   1.50  &  1.00  \\
TrES-3b   & 18.06.20 & STELLA  & B &  88  &   80  &   107  &   1.76  &  1.00  \\
TrES-3b   & 14.07.20 & STELLA  & B & 101  &   80  &   107  &   1.31  &  1.00  \\
TrES-3b   & 22.07.20 & STELLA  & B & 102  &   80  &   106  &   1.34  &  1.00  \\
TrES-3b   & 08.08.20 & STELLA  & B & 102  &   80  &   106  &   1.33  &  1.00  \\  
TrES-3b   & 17.08.20 & STELLA  & B &  89  &   80  &   106  &   1.75  &  1.00  \\
TrES-3b   & 07.07.21 & STELLA  & B &  88  &   80  &   106  &   1.16  &  1.75  \\
TrES-3b   & 16.07.21 & STELLA  & B &  93  &   80  &   107  &   1.60  &  1.09  \\
TrES-3b   & 12.03.22 & STELLA  & B &  89  &   80  &   106  &   1.39  &  1.02  \\
TrES-3b   & 15.04.22 & STELLA  & B & 102  &   80  &   106  &   1.39  &  1.00  \\
TrES-3b   & 19.05.22 & STELLA  & B &  74  &   80  &   106  &   1.20  &  1.00  \\
\hline
\noalign{\smallskip}
TrES-3b   & 06.05.19 & TJO     & V & 116  &   60  &    68  &   1.45  &  1.05  \\
TrES-3b   & 14.05.19 & TJO     & V & 114  &   60  &    68  &   1.86  &  1.00  \\
TrES-3b   & 27.05.19 & STELLA  & V &  69  &   80  &   106  &   0.93  &  1.10  \\
TrES-3b   & 31.05.19 & STELLA  & V & 101  &   80  &   106  &   1.20  &  1.17  \\
TrES-3b   & 17.06.19 & STELLA  & V & 102  &   80  &   106  &   0.83  &  1.00  \\
TrES-3b   & 22.06.19 & STELLA  & V & 100  &   80  &   106  &   0.82  &  1.00  \\
TrES-3b   & 26.06.19 & STELLA  & V & 102  &   80  &   106  &   0.96  &  1.19  \\

\hline

\hline                                                                                                     
\end{tabular}
\end{center}
\end{table*}

\section{Analysis}
\begin{figure}[h!]
\begin{center}
\includegraphics[scale=0.70,angle=270]{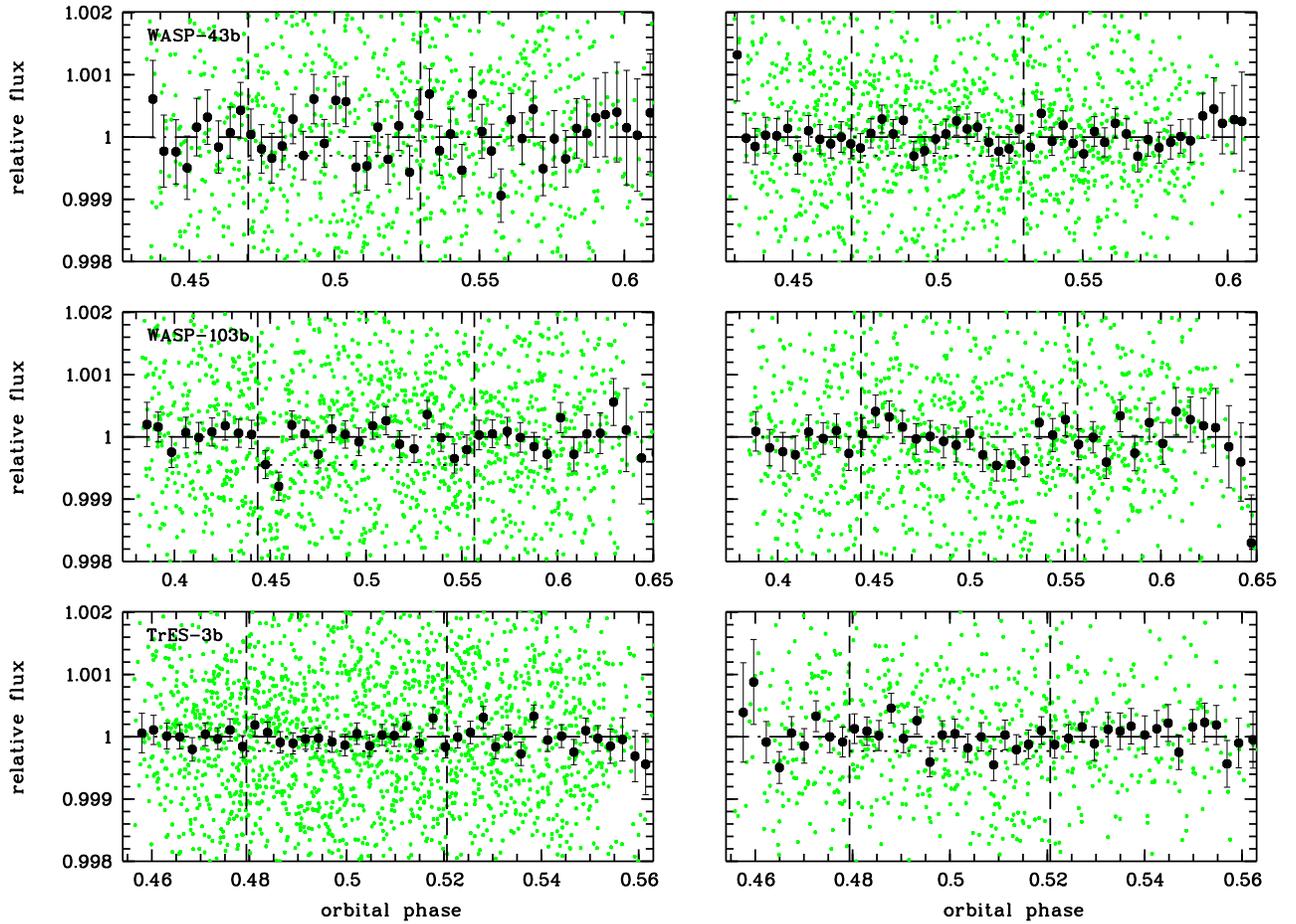}
\caption{Photometric light curves of the three targets of interest, from top to bottom: WASP-43b, WASP-103b, and TrES-3b. The left column shows the data of the Johnson B bandpass, the right column the Johnson V data. In green, the detrended, individually observed data points are shown, in black the detrended, phase-folded, and binned data points. The begin and end of the eclipse event is marked by vertical dashed lines, a zero eclipse depth is marked by a horizontal dash-dotted line, and for visual impression the eclipse depth corresponding to an arbitrary geometric albedo of 0.3 is marked by a horizontal dotted line.   \label{fig:1}}
\end{center}
\end{figure}

The analysis followed closely the detailed description of previous analyses of STELLA exoplanet photometry, e.g. \cite{Mallonn2019,Mallonn2022}. We adjusted the uncertainties of the individual photometric data points per light curve in a two-step process. The light curves were modeled with JKTEBOP \citep{Southworth2005,Southworth2011}. This software tool supplies a transit model, by which we approximate the eclipse depth by the fit parameter of the planet-to-star radius squared $k^2\,=\,(R_p/R_s)^2$. This approach yields inaccurately short ingress and egress durations, but the effect is negligible because our estimation of the eclipse depth is mainly limited by the photometric uncertainty of the in-eclipse and out-of-eclipse data and the uncertainties of the detrending parameters \citep{Mallonn2022}. For simultaneous detrending we used a second-order polynomial in time per light curve \citep{Mallonn2019,Mallonn2022}. All other relevant orbital parameters of the exoplanets were fixed to their literature values. Eclipse detections in the near-infrared have been reported for all three targets \citep{Stevenson2017,Kreidberg2018,Fressin2010}, and the orbits were measured to be nearly circular without timing deviations from a linear orbital ephemeris. Therefore, also the timings of eclipse events were treated as fixed parameters in our analysis. In Figure~\ref{fig:1} we show the detrended, phase-folded and binned light curves of all targets and filters.

\section{Results and discussion}
None of the six data sets, two bandpasses for each of the three targets, yielded a significant detection of the secondary eclipse, despite of the achieved high precision of 90 to 290~ppm depth uncertainty. To test how well the six data sets are able to reveal an astrophysical eclipse signal, we artificially injected in each data set secondary eclipse dips of depth of 1, 2, and 4~ppt. Then we fitted the data sets again with the same setting of free parameters. In all cases, the injected signal of secondary eclipse depth was recovered within $1.5\,\sigma$~uncertainty. The mean of the three values for obtained minus injected eclipse depth is provided here as the final eclipse depth: WASP-43b in Johnson B and Johnson V, respectively: $0.18 \pm 0.29$~ppt and -0.07 $\pm$ 0.11~ppt, WASP-103b in B and V: 0.20 $\pm$ 0.16~ppt and -0.10 $\pm$ 0.18~ppt, and TrES-3b in B and V: 0.08 $\pm$ 0.09~ppt and 0.03 $\pm$ 0.19~ppt. A summary is given in Table~\ref{tab_res}.

These final values for the eclipse depths were transformed to a value of geometric albedo $A_g$ by the standard formulae eclipse depth $d$ equals $A_g (R_p/a)^2$. We derive values of $A_g$ for WASP-43b in B and V of 0.18 $\pm$ 0.29 and 0.00 $\pm$ 0.11, respectively, for WASP-103b in B and V of 0.13 $\pm$ 0.09 and 0.00 $\pm$ 0.12, and for TrES-3b in B and V of 0.11 $\pm$ 0.12 and 0.04 $\pm$ 0.26. These values are not corrected for a contribution of thermally emitted planetary light. Using Eq.~2 of \cite{Mallonn2019}, we estimate such contribution to the eclipse depth of WASP-103b to be $\sim\,50$~ppm and $\sim\,120$~ppm in B and V, respectively, employing literature values for the planetary equilibrium temperature, while for the other two planets it amounts to less than 1~ppm. 

The derived low values of the geometric albedos agree well with reported low values at longer wavelengths for the same planets \citep{Fraine2021,Mallonn2019,Winn2008b}. We find a tentative trend of higher geometric albedos in the filter Johnson B at 445~nm than in Johnson V at 550~nm, however, this trend needs to be confirmed by future observations. In general, this work has proven that meter-sized ground-based telescopes can significantly contribute to measurements of wavelength-resolved reflection properties for selected exoplanet targets.

\begin{table*}
\small
\caption{Overview of the results for secondary eclipse depth and geometric albedo.}
\label{tab_res}
\begin{center}
\begin{tabular}{lcrc}
\hline
\noalign{\smallskip}
Planet    &  Bandpass    & $d$ (ppt) & $A_g$  \\
\noalign{\smallskip}
\hline
\noalign{\smallskip}
WASP-43b  & B & 0.18 $\pm$ 0.29 & 0.18 $\pm$ 0.29 \\
          & V & -0.07 $\pm$ 0.11 & 0.00 $\pm$ 0.11 \\
WASP-103b & B & 0.20 $\pm$ 0.16 & 0.13 $\pm$ 0.09 \\
          & V & -0.10 $\pm$ 0.18 & 0.00 $\pm$ 0.12 \\
TrES-3b   & B & 0.08 $\pm$ 0.09 & 0.11 $\pm$ 0.12 \\
          & V & 0.03 $\pm$ 0.19 & 0.04 $\pm$ 0.26 \\
\noalign{\smallskip}
\hline
\end{tabular}
\end{center}
\end{table*}

\vspace{5mm}
\facilities{STELLA(WiFSIP), TJO(MEIA2) }


\software{JKTEBOP \citep{Southworth2005}
          }



\bibliography{Mallonn_alb}{}
\bibliographystyle{aasjournal}

\end{document}